\documentclass[12pt]{article}
\def\gsim{\, \rlap{$>$}{\lower 1.1ex\hbox{$\sim$}}\,}
\def\lsim{\, \rlap{$<$}{\lower 1.1ex\hbox{$\sim$}}\,}
\usepackage{graphics, color}
\usepackage{graphicx}
\usepackage{amssymb}

\newcommand{\sect}[1]{\section{#1}\setcounter{equation}{0}}

\def\gsim{\, \rlap{$>$}{\lower 1.1ex\hbox{$\sim$}}\,}
\def\lsim{\, \rlap{$<$}{\lower 1.1ex\hbox{$\sim$}}\,}

\begin{document}
\begin{titlepage}
\bigskip
\bigskip\bigskip\bigskip
\centerline{\Large \bf  Cosmic String Loops and Gravitational Radiation}
\bigskip\bigskip\bigskip

\centerline{\large Joseph Polchinski}
\bigskip
\centerline{\em Kavli Institute for Theoretical Physics}
\centerline{\em University of California}
\centerline{\em Santa Barbara, CA 93106-4030} \centerline{\tt joep@kitp.ucsb.edu}
\bigskip


\begin{abstract}
Understanding of the signatures of cosmic string networks is limited by a large uncertainty in the sizes at which cosmic string loops form.  We review cosmic string network evolution, and the gravitational signatures, with emphasis on this uncertainty.  We then review a recent analytic model of cosmic string networks.  In combination with recent simulations, this suggests that 90\% of the string goes into very small loops, at the gravitational radiation scale, and 10\% into loops near the Hubble scale.   We discuss cosmic string signatures in such a scenario, and the `inverse problem' of determining the microscopic cosmic string properties from observations.

To appear in the Proceedings of the 11th Marcel Grossman meeting.
\end{abstract}
\end{titlepage}
\baselineskip = 16pt

\sect{Introduction}\label{intro}

Compactifications of string theory give rise to many potential cosmic strings, including
the fundamental strings themselves, D-strings and wrapped D-branes, solitonic strings and
branes in ten dimensions, and magnetic and electric flux tubes in the four-dimensional effective
theory.  In various models  of inflation in string theory or grand unified theories a network of such strings will form.  Thus it is interesting to ask, what are the prospects for future discovery of these strings if they exist, and, what are the prospects for distinguishing different microscopic models?\footnote{My MG11 talk had substantial overlap with hep-th/0410082~\cite{Polchinski:2004ia}, to which the reader is referred for further details and references.  Here I will concentrate on some recent developments, in particular my work with Jorge Rocha and Florian Dubath.  This is a slightly longer version than in the official proceedings, with the addition of Sec.~3.4.2 and related discussion in Secs.~4.2 and~5.2, and a new title (the title in the proceedings, `The Cosmic String Inverse Problem' seemed too broad in retrospect, since we consider almost exclusively vanilla strings).}

In between the microscopic theories and the observations, the strings have a variety of macroscopic parameters and properties, including:
\begin{itemize}
\item
Their tension, $\mu$.
\item
Their intercommutation probability $P$.  When two strings collide, they either pass through each other, with probability $1-P$, or intercommute (reconnect), with probability $P$.  In fact, $P$ is a function of the velocity and angle of the collision, but to first approximation one can consider the value averaged over the kinematics.
\item
Their gapless degrees of freedom: are these just the collective coordinates for the motion in three-dimensional space, or are there additional bosonic or fermionic modes?
\item
Their interactions with low energy fields: are these gravitational only, or axionic or gauge as well?
\item
The number of kinds of string: one or many?
\item
The existence, or not, of multi-string junctions.
\end{itemize}
One would hope to proceed from the observation of cosmic strings to a determination of their macroscopic properties, and from there to possible microscopic models.\footnote{That is, we are separating the subject into `model building,' the connection between the microscopic and macroscopic models, and `phenomenology,' the connection between macroscopic models and observations.  The present paper (and our references) will be focussed entirely on phenomenology.}
This requires the ability to calculate with some degree of precision the signatures of a given macroscopic model.  In fact, this is quite challenging.  

Here we will focus on the most vanilla type of cosmic string: a single species with $P=1$, no long-ranged interactions besides gravity, and no multi-string junctions or extra degrees of freedom.  Even for these, major uncertainties remain.  Fig.~1 shows a simulation of the network, with one side of the box being about a quarter of the Hubble length.  This simulation is for the radiation era; the matter era simulations have the same general features but some quantitative differences.
\begin{figure}[h]
\begin{center}
\includegraphics[width=23pc]{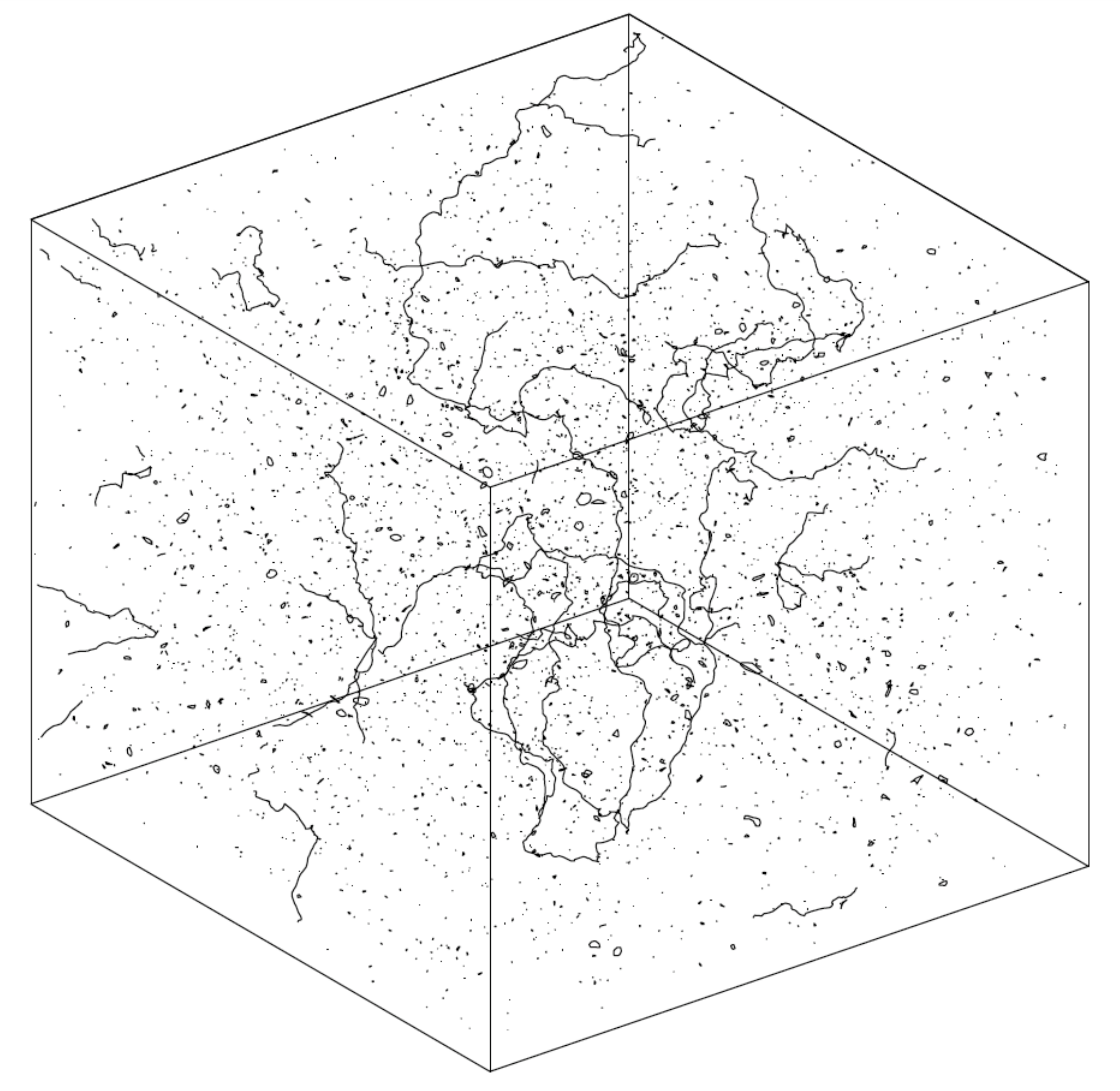}
\end{center}
\caption{Simulation of a vanilla string network during the radiation era, from Allen and Shellard~\cite{Allen:1990tv}.  The scale of the box is approximately one quarter of the Hubble length.}
\end{figure}
One sees several long strings crossing the simulation volume, and a gas of string loops.  

As will be explained in more detail in the next section, the original expectation was that the Hubble length (which is the same order as the horizon length, and the FRW time) would be the only relevant scale in the problem~\cite{Kibble:1976sj}.  If this were true, then simulations, beginning with Refs.~\cite{Allen:1990tv,Albrecht:1989mk,Bennett:1989yp}, should have readily produced good quantitative descriptions of the networks.  However, inspection of Fig.~1 shows that the long strings are not smooth at short distance, but have a kinked structure, which persists down to the cutoff scale of the simulations.  Further, it appears that the loops are rather smaller than the Hubble scale.  Indeed, in the early simulations the loop distribution seemed to peak near the short distance cutoff scale of the simulations.  

Subsequent work has failed to produce a clear picture, and in fact the estimates of the typical size at which the loops are produced range over more than {\it fifty} orders of magnitude.  This is rather remarkable, when one considers that the entire span from the Planck scale to size of the visible universe is just sixty orders of magnitude.  Moreover, the problem is completely well-posed: the limitation is only in our ability to solve the equations.  The problem with the simulations is that they are limited to just a few orders of magnitude in length scale, and even more limited in time scale: for the latter not only is there the direct cost, but the expansion of the universe means that increased time also costs increased simulation volume.\footnote{The horizon length is increasing more rapidly than the comoving length, so the number of degrees of freedom that can interact with one another increases with time.}  Much of the work is based on an expansion factor that is only of order 5, and it is difficult to distinguish transient effects from real ones.

Analytic methods are difficult because of the nonlinearities of the problem.  One can write down the evolution equations for an ensemble of strings~\cite{Austin:1993rg}, but before long one must begin to approximate, and in the end most analytic treatments reduce to somewhat coarse models in terms of a few parameters.  On the other hand, one might have expected that the nature of the problem, the existence of a large ratio of scales, would lend itself to something like a renormalization group treatment.  Roughly speaking, one would use the simulations to treat the strongly nonlinear behavior at the Hubble scale, and then evolve down to shorter scales analytically.

This was the motivation for the recent work of Refs.~\cite{Polchinski:2006ee,Polchinski:2007rg,Dubath:2007wu,wip}.  Thus far, things are not as systematic as for the renormalization group.  The difficulty is evidently that for the string network the flow is not from short distances to long but the other way around.  The Hubble length, against which all things are scaled, increases more rapidly than the comoving length (during the matter and radiation eras), so comoving structures move to smaller effective scales over time.  In this respect the problem is like turbulence, and has been similarly frustrating.

Nevertheless, it has been possible to proceed by making approximations that are not precisely controlled but perhaps well-motivated.  In particular, this has allowed us to understand why structures appear at scales far below the Hubble length.  The simulations are still essential, but the analytic model enables us to distinguish real effects from transient ones, and to extrapolate beyond the scales that can be simulated.  Combined with recent simulations that use a trick to get beyond the limitation on expansion times, it may be that a consistent and fairly quantitative picture is emerging.

In Sec.~2 we review the basic features of network evolution, defining the problem.  In Sec.~3 we discuss some aspects of the gravitational wave signatures, focussing on the sensitivity to loop size.  In Sec.~4 we present the analytic model of the short distance structure.  In Sec.~5 we attempt to compare the model with the simulations, and discuss the current picture.

\sect{Network Evolution}

In this section we review the basics of network evolution, focusing on the issues that will be relevant later.  Aside from some recent work, this is standard material, on which the reader can find more detail in the excellent book by Vilenkin and Shellard~\cite{VilShell}.  

The relevant processes for vanilla strings are:
\begin{enumerate}
\item
The formation of an initial string network, such that a finite fraction of the string goes into infinite random walks.
\item
The stretching of the string network with the expansion of the universe.
\item
The intercommutation of long strings, producing kinks.
\item
The emission of gravitational radiation from long strings.
\item
The self-intercommutation of long strings, producing loops, which may fragment further into smaller loops.
\item
The decay of the loops via emission of gravitational radiation.
\end{enumerate}
We will discuss each of these processes in turn, and then the key issue of the scaling property of the network.

\subsection{The Kibble Process}

Any gauge theory with a broken $U(1)$ symmetry has a string soliton solution.  Moreover, such strings will actually form in any phase transition in which a $U(1)$ symmetry {\it becomes} broken~\cite{Kibble:1976sj}.  At the transition, the Higgs field rolls from the origin to one of the symmetry-breaking vacua, and unless the transition takes place infinitely slowly, causality implies that the phase will uncorrelated over long distances.\footnote{Of course the phase is not gauge invariant, but we can compare it using parallel transport, or in Coulomb gauge.}  Thus defects - strings - will form.  Causality implies that over long distances the strings form random walks.  A random walk has a finite probability not to return to its starting point, so some of the string will be in infinite random walks; this conclusion is supported by simulations.  These infinite strings are the seeds of the later network, because any loops that form in the transition will rapidly decay.

Subsequently, Sarangi and Tye~\cite{Sarangi:2002yt,JST} made the remarkable observation that D-strings would be produced by D-brane annihilation at the end of brane inflation, since the low energy dynamics is again described by $U(1)$ symmetry breaking.  In fact, regardless of the microscopic structure of the strings, there is be some dual version of the Kibble process, such that if the strings can exist only after some phase transition, they will actually form during that transition.  

\subsection*{Stability}
We must assume that the processes of breakage and string confinement~\cite{Witten:1985fp,Preskill:1992ck,Copeland:2003bj} are either absent or sufficiently suppressed so as to be negligible on cosmological time scales.

\subsection{Strings in Expanding Spacetime}

Cosmic strings are long compared to their thickness, and so can be treated as idealized one-dimensional objects.  For vanilla strings, the relevant action is just the Nambu action in the FRW metric
\begin{eqnarray}
ds^2 &=& -dt^2 + a(t)^2 d{\bf x} \cdot d{\bf x} \nonumber\\
&=& a(\tau)^2 (-d\tau^2 + d{\bf x} \cdot d{\bf x})\ .
\end{eqnarray}
The equation of motion governing the evolution of a cosmic string is
\begin{equation}
\ddot{\bf x} + 2\,\frac{\dot{a}}{a}\,(1-\dot{\bf x}^2)\, \dot{\bf x} = \frac{1}{\epsilon} \left( \frac{{\bf x}'}{\epsilon} \right)' \ .
\label{EOM}
\end{equation}
Here $\epsilon$ is given by
\begin{equation}
\epsilon \equiv \left( \frac{{\bf x}'^2}{1-\dot{\bf x}^2} \right)^{1/2}.
\label{epsilon}
\end{equation}
These equations hold in the transverse gauge, where $\dot{\bf x}\cdot{\bf x}'=0$. Dots and primes refer to derivatives relative to the conformal time $\tau$ and the spatial parameter $\sigma$ along the string, respectively. The evolution of the parameter $\epsilon$ follows from equation (\ref{EOM}),
\begin{equation}
\frac{\dot{\epsilon}}{\epsilon} = -2\,\frac{\dot{a}}{a} \, \dot{\bf x}^2.
\label{epsev}
\end{equation}

The equation of motion implies that on scales long compared to the horizon length, which is of the order of the FRW time $t$, the string is frozen in the comoving coordinates $\bf x$, and so just expands with the universe.  When a scale comes inside the horizon, the modes begin to oscillate; the energy begins to redshift away, and the string straightens.

From the second derivative terms it follows that signals on the string propagate to the right and left with $d\sigma = \pm d\tau/\epsilon$.  Thus the structure on a short piece of string at a given time is a superposition of left- and right-moving segments.   In an expanding spacetime the left- and right-moving waves interact --- they are not free as in flat spacetime.  In terms of left- and right-moving unit vectors $\bf{p}_\pm \equiv \dot{\bf{x}} \pm \frac{1}{\epsilon}\bf{x}'$, the equation of motion~(\ref{EOM}) can be written as
\begin{equation}
\dot{\bf p}_\pm \mp \frac{1}{\epsilon} {\bf p}'_\pm = - \frac{\dot{a}}{a} \left[ {\bf p}_\mp - ({\bf p}_+ \cdot {\bf p}_-)\, {\bf p}_\pm \right],
\label{EOM2}
\end{equation}
so that in the limit that we can ignore the time derivative of the scale factor we recover independent left- and right-moving waves.

\subsection{Long String Intercommutation}

When two straight strings intercommute, the resulting strings have a kink, which separates into left- and right-moving kinks.  Under the flat spacetime Nambu equations, the kinks would move indefinitely in the two directions.  The curved spacetime equations imply that the kinks persist but their angle slowly decreases~\cite{Bennett:1989yp}.  The decrease is sufficiently slow that the kinkiness is potentially very large: if one considers the total root-mean-square kink angle in a length of long string (which is what would be relevant if the kink directions were uncorrelated), one finds that it diverges, so there is the potential for having structure on very small scales in the network.

\subsection{Long String Gravitational Radiation}

The oscillations of the long string will emit gravitational radiation, and so are damped below a certain length scale.  A naive estimate would lead to the same scale that we will find below for the loop decays, something of order the dimensionless string tension $G\mu$ times the FRW time $t$.  However, things are more subtle for long strings.  Emission of gravitational radiation requires that a right- and left-moving mode meet.  There is a nonlinear suppression when these have very different wavelengths, because the short-wavelength modes essentially perceives a straight string~\cite{Siemens:2001dx}, so that the actual damping scale is $O(t)$ times a larger power of $G\mu$~\cite{Siemens:2002dj,Polchinski:2007rg}.  We will be more quantitative about this in Secs.~4 and~5.

\subsection{Loop Formation}

When a long string intercommutes with itself, a loop breaks off.  The loop may then self-intercommute and fragment into smaller loops.  This happens rapidly, within one period, so we are interested in the size distribution of the ultimate non-self-intersecting loops.
Of course a loop may collide with a long string (or each other) and reattach; for loops close to the FRW scale this is likely, but smaller loops have a high probability never to rejoin.

\subsection{Loop Decay}

Dimensionally, a loop of length $l$ emits gravitational radiation power of order $G \mu^2 $.  Since its total energy is of order $\mu l$, we conclude that the lifetime is
\begin{equation}
t(l) = l/\Gamma G \mu \ ,
\end{equation}
where the numerical factor $\Gamma$ is found to be of order 50 for typical loops.  In other words, loops smaller than $\Gamma G \mu t$ will decay in a Hubble time.

\subsection{Scaling and Loop Size}

Now let us put the preceding ingredients together.  A key concept is {\it scaling,} the idea that the statistical properties of the network should be constant when measured in units of the FRW time $t$~\cite{Kibble:1976sj}.  For example, the mean separation between long strings should grow as $t$.  If the network simply expanded with the universe, this separation would grow as the scale factor $a(t)$, which increases more slowly than $t$.\footnote{Note that this discussion is relevant only during the periods of matter and radiation domination.  We will discuss some effects of cosmic acceleration later.}  Scaling requires that the processes discussed above reduce the amount of string, and that they do so as fast as causality allows, since the horizon distance is also linear in $t$.  For a network that scales, the energy in strings in a box of side $t$ is linear in $t$, so the density is proportional to $t/t^3 = 1/t^2$.  Thus it is a fixed multiple of the dominant energy in each epoch, which also is proportional to $H^2 \propto t^{-2}$.   If the energy density did not scale, the strings would eventually come to dominate, unless the string tension were very low.

In fact, it appears that string networks scale under rather general conditions, so that the scaling solution is an attractor.  If the string density exceeds the attractor value, the rate of long-string collisions increases, resulting in more kinks and more loop formation; if it is smaller, than the opposite occurs.  A simple {\it one-scale model} captures this behavior~\cite{Kibble:1984hp}.  Suppose that there is a single length scale $L$ that governs the network, that is, the long-string separation, the correlation of structure along the long string, and the  formation of loops.  A convenient definition is in terms of the energy density in long strings, $\rho_\infty = \mu/L^2$.  The energy of long strings in a comoving volume would then satisfy
\begin{equation}
\frac{dE_\infty}{dt} = \frac{1}{a}\frac{d a}{dt} ( 1 - 2\bar v^2 ) E_\infty - \frac{c}{L}  E_\infty \ .
\label{1sm}
\end{equation}
The first term is from the expansion of the universe: the 1 in parentheses is just the growth of the length with the scale factor, while the $-2 \bar v^2$ is from the redshifting away of energy below the horizon scale.  The latter follows from the definition~(\ref{epsilon}), which implies that the total energy in a string is simply $\mu \int d\sigma\, a\epsilon$, together with the equation of motion~(\ref{epsev}).  
The second term is from loop formation, with $c$ a constant and the scale $L$ inserted by dimensional analysis.

The average $\bar v^2$ plays an important role in the properties of the network.  If the universe were not expanding, the virial theorem would imply that  $\bar v^2 = 0.5$.  The expansion of the universe reduces this to around $0.41$ in the radiation era and $0.35$ in the matter era, according to the simulations~\cite{Martins:2005es}.

Inserting $E_\infty = a^3 \mu /L^2$, and defining $x= L/t$ and $a \propto t^r$, one finds that
\begin{equation}
\frac{t}{x} \frac{dx}{dt} = (1 + \bar v^2) r  - 1 + \frac{c}{2x}\ .
\end{equation}
Since $\bar v^2$ is less than 0.5, the first term is less than 1 in both the matter and the radiation eras, and so $x$ is driven to a fixed point value $c/2(1 - [1 + \bar v^2]r)$.  In fact this model is overly crude and has been superseded by more elaborate models with multiple scales and other parameters, but it captures the basic feedback that is operating.

The scaling property means that the initial conditions from the formation of the network, in particular the initial density of strings, are washed out in time.  Thus there are only three processes that enter: expansion, intercommutation, and gravitational radiation.  However, there are some important questions that remain open:  Does the scaling property apply only to the gross features of the network, or to all details including the loop distribution?  And, does gravitational radiation only play the role of removing the loops from the network, or is gravitational smoothing of long strings important as well?  Note that if long-string gravitational radiation can be ignored, then the problem has no parameters: the processes of expansion, intercommutation, and loop removal are purely geometric (the string tension $\mu$ scales out of the equations of motion, and $P$ has been set to 1).

As we have noted in the introduction, there is still uncertainty about the small scale properties of the network.  For example, the estimates of the characteristic size $l_{\rm f}$ at which loops form range over
\begin{itemize}
\item
$l_{\rm f} \sim 0.1 t$.  A loop size near the Hubble scale was the initial expectation in this subject, as embodied in the one-scale model, and in fact this is also the conclusion of some of the most recent simulations~\cite{Vanchurin:2005pa,Olum:2006ix}.
\item
$l_{\rm f} \sim 10^{-2}$ to $10^{-3} t$.  Other recent simulations~\cite{Ringeval:2005kr,Martins:2005es} point to a loop size that still scales as a pure number times the FRW time, but rather smaller than the above.
\item
$l_{\rm f} \sim \Gamma G \mu t$.  These loops still scale, but only with the inclusion of gravitational backreaction to smooth the long strings~\cite{Bennett:1989ak}.
\item
$l_{\rm f} \sim \Gamma (G \mu)^{k > 1 } t$.  Again this length scales, but as noted earlier backreaction is less efficient than originally thought~\cite{Siemens:2001dx,Siemens:2002dj,Polchinski:2007rg}.
\item
$l_{\rm f} \sim $ string thickness.  This length does not scale, but remains at a fixed physical size~\cite{Vincent:1996rb,Vincent:1997cx,Bevis:2007gh}.
\end{itemize} 
This has been a contentious subject, and indeed in the references a number of names are attached to more than one estimate.  In the end this will not be surprising, as we will argue that more than one of these estimates is correct.

\sect{Network Signatures}

Since vanilla strings have only gravitational long-range interactions, the key parameter is $G\mu$, which determines the typical metric perturbation produced by a long string.  In models of brane inflation, this is related to the scale of CMB fluctuations $\delta T / T$, but in a way that depends on the geometry so that one obtains as an interesting range~\cite{JST}
\begin{equation}
10^{-12} < G\mu < 10^{-6}\ .
\end{equation}
This is probably also relevant to more general models in which strings are produced at the end of inflation.  Possible gravitational signatures include
\begin{itemize}
\item
Dark matter.
\item
Effect on CMB and galaxy formation.
\item
Gravitational lensing.
\item
Gravitational radiation.
\end{itemize} 

\subsection{Dark matter}

We can immediately eliminate this.  We have noted that the string density is a fixed fraction of the dominant energy density in a scaling regime.  The ratio is proportional to the string tension, in the dimensionless combination $G\mu$;
the simulations give approximately $70 G\mu$ in the matter era and $420 G\mu$ in the radiation era.  Since the current upper bound on $G\mu$ is around $2 \times 10^{-7}$, this is rather less than one.  Of course, if the network does not scale and if the string tension is just right, then the strings could become dominant at some late time  --- but then they would not redshift like matter.   The effective value of the equation of state parameter would be between $0$ and $-1/3$, so they are also not the dark energy.

\subsection{CMB and Galaxies}

The CMB perturbations come primarily from the long strings, since their fields extend coherently over long distances.  Thus they depend on network properties that are reasonably well understood.\footnote{However, Ref.~\cite{Bevis:2007gh} finds densities several times lower, so the bounds are several times weaker.  It also rather optimistically suggests that a best fit per degree of freedom includes cosmic strings.}  The scaling property implies that this perturbation spectrum is scale invariant, but the power spectrum does not match observations: there are no acoustic peaks, because the perturbations are generated after the modes reenter the horizon rather than (as with inflation) before they leave.  This translates into an upper bound~\cite{Pogosian:2006hg,SelSlo} $G\mu \lsim 2 \times 10^{-7}$.  Bounds from non-gaussianities are currently less sensitive than those from the power spectrum.  The strongest bound, around $3 \times 10^{-7}$ comes not from limits on string-like features, but from limits on Doppler deviations from a black body spectrum that would be introduced by the transverse motion of the strings~\cite{Jeong:2004ut}. The power spectrum bound is argued to be approaching the cosmic variance limit, but bounds from polarization~\cite{Pogosian:2006hg,SelSlo} or pattern search will improve one or two orders of magnitude with time.

If the loops are large enough, they can act as seeds for early galaxy formation and so lead to early reioniziation~\cite{Olum:2006at}.  If the largest estimates of loop size are correct, this leads to a stronger bound $3 \times 10^{-8}$, with some uncertainties.

\subsection{Lensing}

Both long strings and loops (if they are large enough) can lens more distant objects.  The deficit angle at a string is $8\pi G\mu$, or around $1$ arc-second at the current CMB bound.  The separation for a given object will be somewhat larger or smaller than this, depending on the geometry and the string velocity~\cite{Shlaer:2005gk}.  Occasional lens candidates have been reported, but have thus far always proven to be binary galaxies.  To obtain bounds from lensing requires a large survey, because a string network lenses only a small fraction of the sky.  Optical surveys may reach $G\mu \sim 10^{-8}$~\cite{Gtap}, and
radio frequency surveys may eventually be sensitive down to $G\mu \sim 10^{-9}$~\cite{Mack:2007ae}.

For lensing, there is the issue of how straight the string is at short distances.  Most phenomenology has considered two extremes: a straight string, which simply translates a strip of the background and so produces identical double images, or a random walk, which produces complex multiple images.  Further, with the straight string any additional lensed objects would be on a line through the lens axis; with the random walk they will be distributed more irregularly.  It will be one of our results to resolve this for the real networks.

\subsection{Gravitational Radiation}

This seems the most likely route by which cosmic strings will be discovered, or the bounds tightened to an interesting degree.  Again, we refer the reader to Vilenkin and Shellard~\cite{VilShell} for a detailed treatment, as well as Ref.~\cite{Allen:1996vm}, and also Refs.~\cite{Hogan:2006we,Siemens:2006yp,DePies:2007bm} for recent updates.  
In contrast to the signatures above, the gravitational radiation comes predominantly from the loops, because these are smaller and so have higher frequencies.  Thus the issue of loop size is particularly important.

For vanilla strings, all of the energy going into loops ends up as gravitational radiation.  We can get the total energy going into loops from energy conservation.  As in Sec.~2.7, we have
\begin{equation}
\frac{d\rho_\infty}{dt} = -\frac{2}{t} [ 1 + \bar v^2] r   \rho_\infty - \frac{d\rho_{\rm loop}}{dt} \ .
\end{equation}
Putting in scaling, $\rho_\infty \propto t^{-2}$, gives 
\begin{equation}
\frac{d\rho_{\rm loop}}{dt} = \frac{2}{t} ( 1 - [1 + \bar v^2] r  ) \rho_\infty  \ . \label{rholoop}
\end{equation}
There are two key distinctions to make: between the radiation from the low and high harmonics on a loop,
and between loops that are large or small compared to $\Gamma G\mu t$.  Smaller loops decay essentially at once on the cosmological time scale, so this energy goes directly into gravitational waves.  Loops that form at larger sizes persist for a while before decaying.  We consider the latter case first.

\subsubsection{Large Loops, $\alpha > \Gamma G \mu$}

Once a loop forms (below the Hubble scale), it stops redshifting, so the energy density in loops behaves like matter.  During the matter era, this is not so interesting, but during the radiation era it implies that the energy density in a set of string loops grows relative to the dominant energy $\rho_{\rm r}$.  A large loop that forms at size $\alpha t_{\rm f}$ will decay at $t_{\rm d}$, where $\alpha t_{\rm f} = \Gamma G\mu t_{\rm d}$.  Thus there is a relative enhancement by $a(t_{\rm d})/a(t_{\rm f}) = (\alpha/ \Gamma G\mu)^{1/2}$, if the loop both forms and decays during the radiation era.  For these loops, then,
\begin{equation}
t_{\rm d} \frac{d\rho_{\rm GW}}{dt_{\rm d}} = 2 ( 1 - [1 + \bar v_r^2] r_{\rm r}  ) \frac{2}{3} \gamma^{-3/2} (\alpha/ \Gamma G\mu)^{1/2} \frac{\rho_\infty}{\rho_{\rm r}} \rho_{\rm r}\ .
\end{equation}
We have included two additional factors found in a more complete treatment: a $\frac{2}{3}$ from integrating the $a(t_{\rm d})/a(t)$ over the life of the loop, and a factor of $\gamma^{-3/2}$ where $\gamma$ is the Lorentz factor with which the loop is produced.  A factor of $\gamma^{-1}$ enters directly because the kinetic energy redshifts away rather than becoming gravitational radiation, and an additional $\gamma^{-1/2}$ enters through $a(t_{\rm d})/a(t_{\rm f})$ because this reduces the lifetime of the loop.

A loop of length $l$ has period $l/2$ and fundamental frequency $\nu = 2/l$.  Most of the energy goes into the lowest few harmonics, so
$\nu(t_{\rm d}) \sim 2/\Gamma G\mu t_{\rm d}$.  The observed frequency is
\begin{equation}
\nu_0 = \frac{a(t_{\rm d}) }{a_0} \nu(t_{\rm d})  = \frac{a(t_{\rm d}) }{a_0}\frac{ 2 }{\Gamma G\mu t_{\rm d}}\ .  \label{nu0}
\end{equation}
The condition that this be during the radiation era is $G\mu  >  1.5 \times 10^{-10} / (\nu_0 \cdot {\rm yr})$.  For LIGO $(\nu_0 \sim 10^2$ Hz) and LISA $(\nu_0 \sim 10^{-2}$ Hz) this inequality is satisfied down to very small tensions.  For pulsar timing ($\nu_0 \sim 0.1$/yr) it is satisfied down to $G\mu$ around $10^{-9}$ (for smaller values of $G\mu$, the gravitational radiation in the pulsar range first rises and then falls steeply).

Finally, noting that during the radiation era $d\nu_0/\nu_0 = -dt_{\rm d}/2t_{\rm d}$, and that the gravitational wave energy redshifts like radiation, we have
\begin{eqnarray}
\nu_0 \frac{d\Omega_{\rm GW}}{d\nu_0} &=&  4 ( 1 - [1 + \bar v_{\rm r}^2] r_{\rm r}  ) \frac{2}{3} \gamma^{-3/2} (\alpha/ \Gamma G\mu)^{1/2} \frac{\rho_\infty}{\rho_{\rm r}} \Omega_{\rm r} 
\nonumber\\
&=& 4 ( 1 - [1 + \bar v_{\rm r}^2] r_{\rm r}  ) \frac{2}{3} \gamma^{-3/2} (\alpha/ \Gamma G\mu)^{1/2} \frac{\rho_\infty}{\rho_{\rm r}} (1 + z_{\rm eq})^{-1} \Omega_{\rm m}
\nonumber\\
&=&  0.0035 \gamma^{-3/2} (\alpha G\mu)^{1/2} \ . \label{stoch}
\end{eqnarray}
In the last line we have inserted $\bar v_{\rm r}^2 = 0.41$, $r_{\rm r} = \frac{1}{2}$, ${\rho_\infty}/{\rho_{\rm r}} = 420 G\mu$, $z_{\rm eq} = 3200$, and $\Omega_{\rm m} = 0.24$.
Note the factor of $\Omega_{\rm m}$: the fractional energy density in gravitational waves is diluted by the vacuum energy: this is a suppression relative to older estimates.

The observed regularity of pulsar signals puts limits on the spacetime fluctuations through which they pass, such that~\cite{jenet} $\nu_0\, {d\Omega_{\rm GW}}/{d\nu_0} < 4 \times 10^{-8}$ for $\nu_0 \sim (10 \,{\rm yr})^{-1}$.  Thus
\begin{equation}
G\mu < 1.3 \times 10^{-10} \alpha^{-1} \gamma^{3}\ . \label{pulbound}
\end{equation}
If we take the largest estimate, $\alpha \sim 0.1$, and assume that $\gamma \approx 1$, we get the strong limit $G\mu < 1.3 \times 10^{-9}$.  The pulsar bounds are likely to improve substantially with time.  Further, Advanced LIGO is expected to reach values of $\nu_0\, {d\Omega_{\rm GW}}/{d\nu_0}$ between $10^{-8}$ and $10^{-9}$, and LISA will reach $10^{-12}$ or better.  This makes clear the importance of understanding the loops.

The emission of gravitational radiation from the high harmonics of the loops is larger than it would be if the string were smooth.  This is first of all due to the kinks, which give a $\nu^{-5/3}$ spectrum ($\nu^{-2/3}$ per log frequency), but an even more interesting effect is due to the cusps.  The functions~${\bf p}_\pm$ represent two curves on the unit sphere, which will generically intersect at points~\cite{Turok:1984cn}.  Each intersection represents a periodic event, in with the string snaps like a whip whose tip approaches the speed of light.  This produces a cone of gravitational waves, with a power spectrum $\nu^{-4/3}$  ($\nu^{-1/3}$ per log frequency) and a distinctive signature in time.

Initial calculations of this signal~\cite{Damour:2001bk,Damour:2004kw} indicated a good chance that it would be seen at LIGO, but more recent calculations~\cite{Siemens:2006vk}, which treat the cosmology and the detector characteristics more precisely, indicate that we may have to wait for LISA.  The cusp signal is less sensitive to the loop size (as long as it is at least $\Gamma G\mu t$) because over much of parameter space one is seeing cusps from the matter era.  However, the largest loop size, $l \sim 0.1 t$, gives a signal within the reach of Advanced LIGO down to $G\mu \sim 10^{-9}$~\cite{Siemens:2006vk}.

\subsubsection{Small Loops}

It is customary to write the formation sizes of these in terms of a parameter $\epsilon < 1$, where $l_{\rm f} = \epsilon\Gamma G\mu t$.  For these, the decay is essentially instantaneous and so the rate of gravitational wave production is
\begin{equation}
t \frac{d\rho_{\rm GW}}{dt} = 2 ( 1 - [1 + \bar v_m^2] r_{\rm m}  ) {\rho_\infty}\ .
\end{equation}
For the low frequency loops, we have 
\begin{equation}
\nu_0 = \frac{a(t_{\rm d}) }{a_0}\frac{ 2 }{\epsilon\Gamma G\mu t_{\rm d}}\ .
\end{equation}
Observations at a given frequency $\nu_0$ are therefore sensitive only to the parameter range
\begin{equation}
\epsilon G\mu \gsim \frac{ 2 }{\Gamma  \nu_0 t_0}\ . \label{mingmu}
\end{equation}

Noting that $\Omega_{\rm GW}$ will be diluted by a factor of $\Omega_{\rm m} a(t_{\rm d}) /{a_0}$,\footnote{For loops that form during the accelerating era one needs to redo the network models, but dilution by $\Omega_{\rm m}$ should be a good estimate.} we have for the low-harmonic radiation
\begin{equation}
\nu_0 \frac{d\Omega_{\rm GW}}{d\nu_0} = 6 ( 1 - [1 + \bar v_{\rm m}^2] r_{\rm m}  ) \frac{a(t_{\rm d}) }{a_0} \frac{\rho_\infty}{\rho_{\rm m}} \Omega_{\rm m} \ . \label{loharm}
\end{equation}
On the other hand, for high-harmonic radiation (log-frequency spectrum $\nu^{-1/3}$) emitted in the very recent past we have
\begin{equation}
\nu_0 \frac{d\Omega_{\rm GW}}{d\nu_0} = 6 c ( 1 - [1 + \bar v_{\rm m}^2] r_{\rm m}  ) 
(  2 / \epsilon \Gamma  G\mu \nu_0 t_0)^{1/3}\
 \frac{\rho_\infty}{\rho_{\rm m}} \Omega_{\rm m} \ . \label{hiharm}
\end{equation}
Here $c$ is a possible suppression factor if many loops do not have cusps.  In Ref.~\cite{Dubath:2007wu} it is argued that small loops are likely to have cusps, so this factor is probably close to 1.
In the range~(\ref{mingmu}), for $c$ of order 1, the high-harmonic signal~(\ref{hiharm}) is greater than the low-harmonic signal~(\ref{loharm}): the suppression of the latter due to redshifting is significantly greater than the suppression of the former from the falloff of the spectrum (and this would still be true even if the high-frequency spectrum were only due to kinks).
Inserting the appropriate constants, Eq.~(\ref{hiharm}) becomes
\begin{equation}
\nu_0 \frac{d\Omega_{\rm GW}}{d\nu_0} = 3.5 c 
 ( \epsilon\nu_0 t_0)^{-1/3}
(G\mu)^{2/3} \ . \label{hiharm2}
\end{equation}
The bounds~(\ref{mingmu},$\,$\ref{hiharm2}) are in approximate agreement with the sensitivity found in Ref.~\cite{Siemens:2006yp} (their Fig.~2, extrapolated to $P = 1$).

Note that Eq.~(\ref{hiharm2}) represents the contribution of the cusps when they are averaged out into a stochastic signal.  In Sec.~5.2 we will discuss the observation of individual cusps.

\sect{A Model of Short Distance Structure}
\subsection{The Long-String Two-Point Function}

We now attempt to get an analytic understanding of the small-scale structure in the network~\cite{Polchinski:2006ee,Polchinski:2007rg}.
  Thus we focus on the evolution of a short left- or right-moving segment on a long string.  This will potentially involve the following:
\begin{enumerate}
\item
Evolution according to the FRW-Nambu equations Eqs.~(\ref{EOM2}).
\item
Long string intercommutation.
\item
Incorporation into a loop larger than the segment.
\item
Emission of a loop comparable to or smaller than the segment.
\item
Smoothing via gravitational radiation.
\end{enumerate}
The probability of the second of these is proportional to the length of the segment, and so can be systematically neglected for a short segment.  The third process is governed by dynamics on a longer distance scale, and so does not depend directly on the configuration of the segment: it will not change the statistical distribution for configurations of small segments.  

In saying this we are neglecting the correlation between the structure at short distance and the structure at long distance.  This is not a fully controlled approximation, but should become progressively better at shorter scales.  One could attempt to make an improved model in which this correlation is parametrized, but this will not be necessary for our purposes.  Ultimately one would like to write down the exact equations for an ensemble of string networks, and then approximate systematically, but this is apt to be very difficult; it is more practical to first figure out what is going on.  

The small loop production would be absent if the only scale were the Hubble length.  We will see that if we assume this and proceed, it will not be self-consistent.  However, we will also find that the production of small loops is primarily controlled by the long-distance structure rather than the short distance structure, so we seem to get the right answer anyway.  Finally, we will assume that, while we are below the Hubble scale, we are at large enough scales to ignore gravitational radiation.

Thus we need only to solve for the motion of a string in an expanding universe.  We recall the equations for the unit vectors $\bf{p}_\pm \equiv \dot{\bf{x}} \pm \frac{1}{\epsilon}\bf{x}'$,
\begin{equation}
\dot{\bf p}_\pm \mp \frac{1}{\epsilon} {\bf p}'_\pm = - \frac{\dot{a}}{a} \left[ {\bf p}_\mp - ({\bf p}_+ \cdot {\bf p}_-)\, {\bf p}_\pm \right]\ .
\end{equation}
These equations
are nonlinear and not solvable in closed form, but again we can simplify for a short segment.  We separate the configuration into a mean direction of the segment and a fluctuation,
\begin{equation}
{\bf p}_\pm(\tau,\sigma) = {\bf P}_\pm(\tau) + {\bf w}_\pm(\tau,\sigma)
-\frac{1}{2}  {\bf P}_\pm(\tau) {w}^2_\pm(\tau,\sigma) + \ldots\ ,
\label{ppw}
\end{equation}
where $P_{\pm}^2 = 1$ and ${\bf P}_\pm \cdot {\bf w}_\pm = 0$.
We will expand in powers of ${\bf w}_\pm(\tau,\sigma)$: the deviation from the mean decreases with the size of the segment (one can show this from the full equations, and we will confirm that it is self-consistent).  Then one finds
\begin{eqnarray}
\dot{\bf P}_+ &=& - \frac{\dot{a}}{a} \left[ {\bf P}_- - ({\bf P}_+ \cdot {\bf P}_-)\, {\bf P}_+ \right]\ ,
\\
\dot{\bf w}_+ - \frac{1}{\epsilon} {\bf w}'_+ 
&=& - ({\bf w}_+ \cdot \dot{\bf P}_+)\, {\bf P}_+ +  \frac{\dot{a}}{a} ({\bf P}_+ \cdot {\bf P}_-)\, {\bf w}_+\ .
\label{EOM3}
\end{eqnarray}
In Eq.~(\ref{EOM3}) we have dropped a term $\frac{\dot{a}}{a} ({\bf P}_+ \cdot {\bf w}_-)\, {\bf P}_+$, because it averages to zero (with corrections suppressed by the length of the segment) as the left- and right-moving waves sweep past each other.  The first term on the RHS of Eq.~(\ref{EOM3}) is just a precession, which keeps ${\bf w}_+$ perpendicular to ${\bf P}_+$; if we work with parallel transported axes we can ignore it.  In the second term, the factor of $\dot a/a$ means that the variation is significant only over a Hubble time.  We then replace ${\bf P}_+ \cdot {\bf P}_-$ with its ensemble-averaged value, which is simply $2\bar v^2 - 1$.  Again, this is not a fully controlled approximation, but should become progressively better for shorter segments.

Along a left-moving characteristic we then have ${\bf w}_+ \propto a^{2\bar v^2 -1} \propto t^{r(2\bar v^2 -1)}$.  Here we take $a = t^r$, so that $\tau \propto t^{1-r}$.  Averaging over a translation-invariant ensemble of segments gives
\begin{equation}
\left\langle [{\bf w}_+ (\sigma,\tau) - {\bf w}_+(\sigma',\tau)]^2 \right\rangle
= t^{-2r(1 - 2\bar v^2 )} f(\sigma-\sigma')\ .
\label{ww}
\end{equation}
The physical length $l$ of a segment of coordinate length $\delta \sigma$ is 
\begin{equation}
l = a(t) \epsilon(t) \delta \sigma \sim a(t)^{1 - 2\bar v^2} \delta \sigma \ , \label{physl}
\end{equation}
 where we have averaged in the $\epsilon$ equation of motion~(\ref{epsev}).\footnote{To be precise, from the definition~(\ref{epsilon}) of $\epsilon$ it follows that $l$ is the energy of the segment divided by $\mu$.  This is the length at an instant when the string is at rest, but it is convenient to work with this energy-length because it is conserved as the string oscillates.  \label{lfoot}}
  As time goes on, the physical length of the segment increases as $a^{1 - 2\bar v^2}$, but this is much slower than the growth of the Hubble length $O(t)$, so the segment `propagates' to shorter and shorter scales.  Correspondingly, as we evolve back, there will be a point at which its length approaches $t$ and our approximations break down.  At this point, we have to let the simulations deal with the horizon-scale dynamics, providing an initial condition that determine the constant $t$.  That is, the correlator~(\ref{ww}) takes some value when $a^{1 - 2\bar v^2} \delta \sigma = Ct$, where the constant $C$ defines the matching scale; i.e. $\delta \sigma = Ct^{1 + r(2\bar v^2 - 1)}$.  This constant will be independent of time if the Hubble-length dynamics scales, so we can conclude that
\begin{equation}
f(\sigma - \sigma') = 2A  |\sigma-\sigma'|^{2 \chi}\ ,\quad
\chi = \frac{ r (1 - 2\bar v^2 ) }{ 1 - r(1 - 2\bar v^2)}
\end{equation}
to cancel the
 time-dependence of the RHS of Eq.~(\ref{ww}) at the matching point. 
In all,
\begin{equation}
\left\langle [{\bf w}_+ (\sigma,\tau) - {\bf w}_+(\sigma',\tau)]^2 \right\rangle
= 2{\cal A} (l/t)^{2\chi} \ .
\label{fluct}
\end{equation}

Thus we obtain a definite form for the two-point function.  Only a constant $\cal A$ need be taken from the simulations, aside from the ubiquitous $\bar v^2$.  In particular, the latter determines the exponents $\chi_{\rm m} = 0.25$ and $\chi_{\rm r} = 0.10$.  We compare the calculated two-point function with simulations~\cite{Martins:2005es} in Fig.~2. 
\begin{figure}
\vspace{-1in}

\begin{center}
\includegraphics[width=30pc]{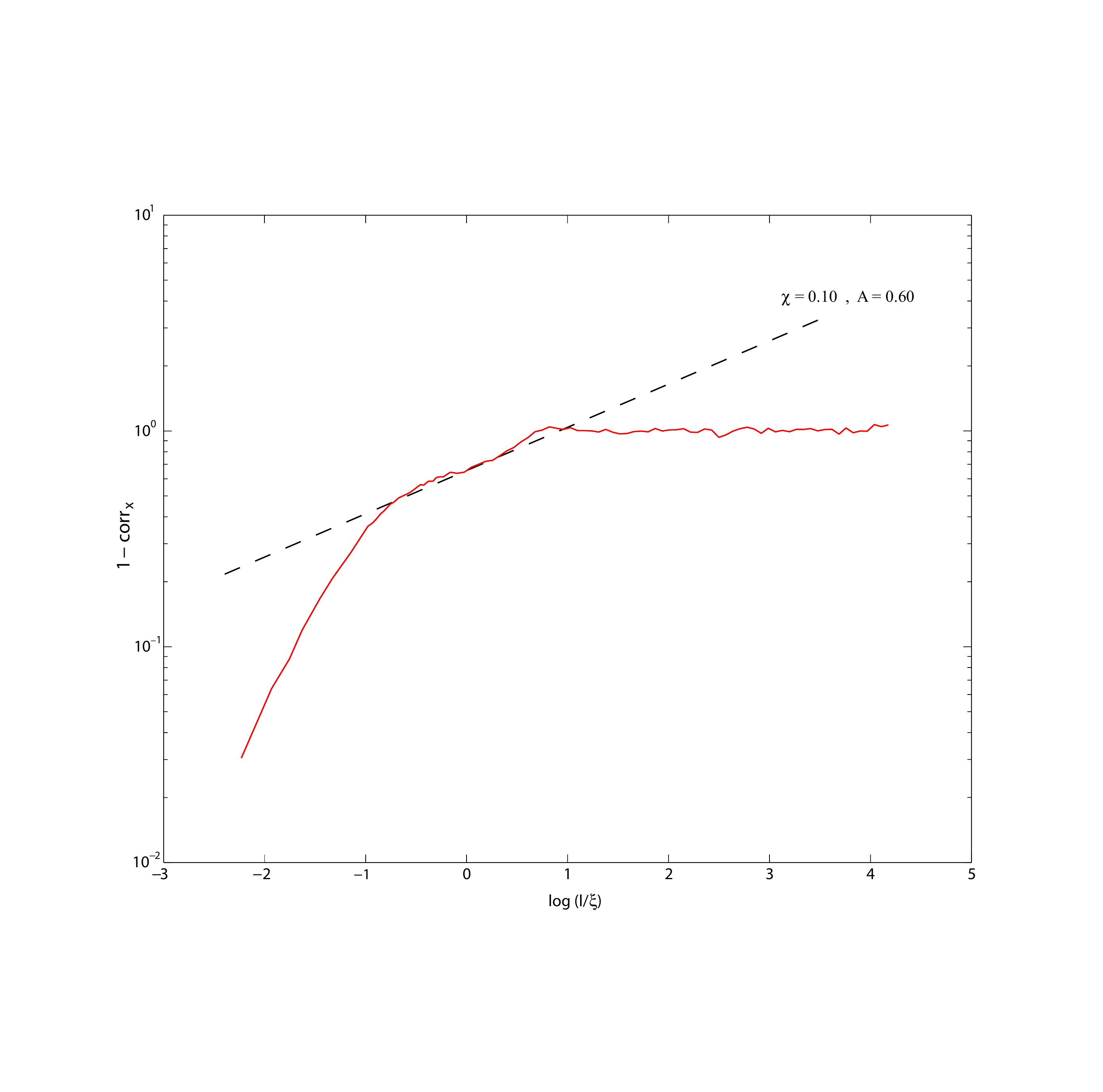}
\vspace{-1.4in}

\includegraphics[width=30pc]{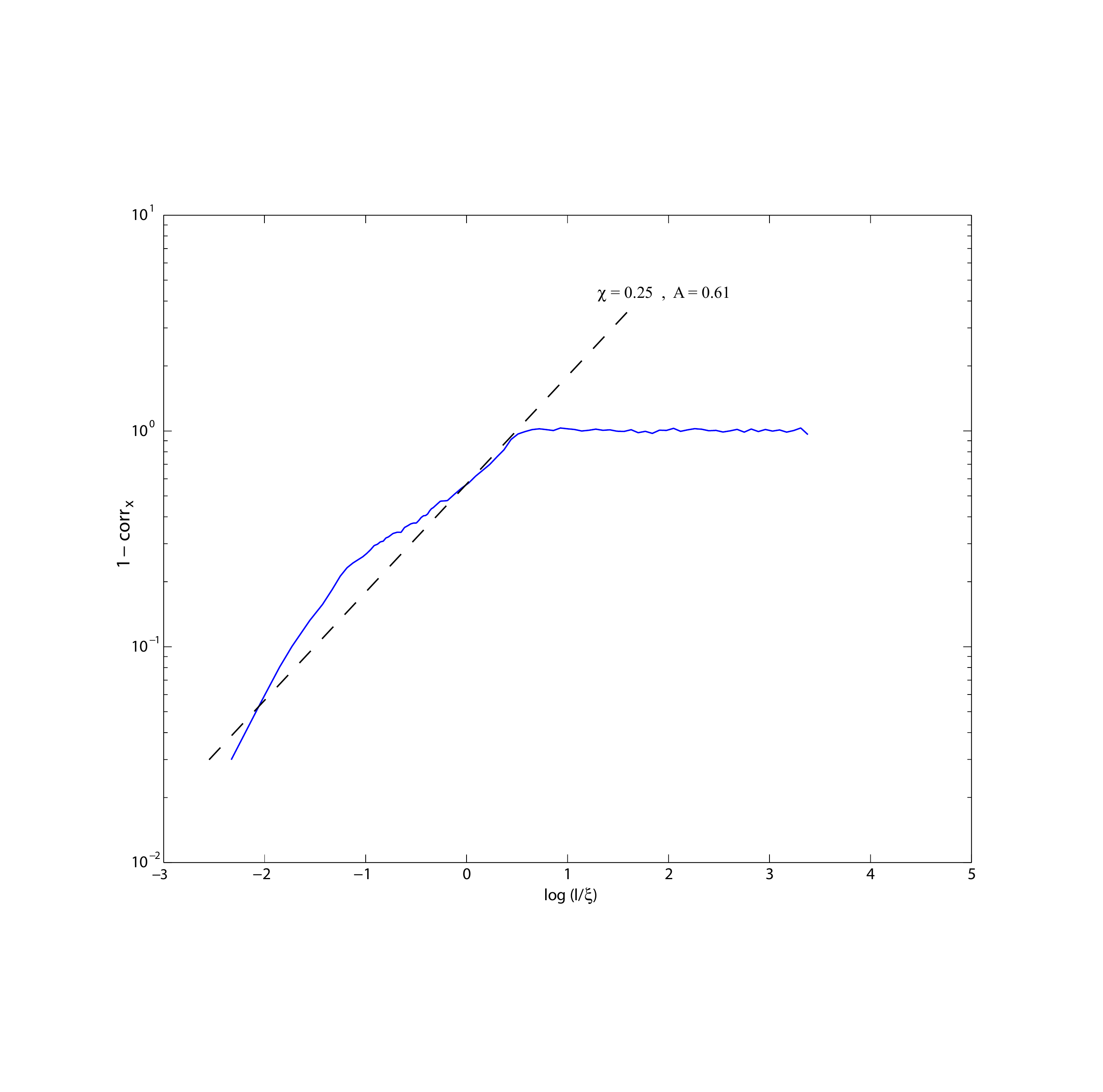}
\vspace{-.7in}

\caption{Log-log comparison of the model (dashed line) with the data provided by Ref.~\cite{Martins:2005es} in the radiation-dominated era (upper plot) and 
the matter-dominated era (lower plot).   The $x$-axis is plotted in terms of a correlation length $\xi$, which is $0.30 t$ in the radiation era and $0.69t$ in the matter era.}
\end{center}
\end{figure}
The log-log slope agrees rather well with the predicted value for a range of scales around near $t$, with ${\cal A}$ around $0.60$ in both eras.  At longer distance it goes over rather abruptly to zero slope, corresponding to the expected random walk.  At shorter distances the simulations and the model differ.  This is not a surprise.  Our picture is that the structure propagates from the Hubble scale down to smaller scales.  The expansion factor for the simulations~\cite{Martins:2005es} is only of order 3, so only a half-order of magnitude could have reached its asymptotic form; at shorter distances it must still reflect in part the initial conditions.\footnote{
We note that the authors of Ref.~\cite{Martins:2005es} report that the same two-point function is obtained with other initial conditions, but this has not been quantified: it would be interesting in particular to compare the time-dependence with our form~(\ref{ww}).} 
  In fact, the agreement is better than might be expected.\footnote{There is an adage (Fermi?) that when one gets the physics right, one's approximations work better than they should.}  After a very long period of time, we would expect the power law to extend to arbitrarily short scales.  Our exponents may be somewhat off, but the asymptotic power law form seems likely to be robust until new physics, particularly gravitational radiation, enters.
  
The correlator of the tangent vector can be written in terms of Eq.~(\ref{fluct}), and the fact that this goes to zero with $l$ means that the string become straighter at short distance.  In other words, the fractal dimension goes to 1,\footnote{This was also to conclusion of Ref.~\cite{Martins:2005es}; the log-log plot accentuates the differences between our results and theirs, whereas the graphs of the fractal dimensions look more similar.}
as $1 + O([l/t]^{2\chi})$.  Thus we can answer the question that arose in the lensing discussion: long strings produce nice double images, with corrections~\cite{Polchinski:2006ee} of order 1\%.  More complex images would be produced if we were near a (rare) large kink, or for lensing by a small enough loop.

Most attempts to study cosmic string networks analytically have worked with a few macroscopic parameters, along the lines of the one-scale model.  Ref.~\cite{Austin:1993rg} is the most systematic attempt to study the evolution of an ensemble analytically, but they did not attempt to explore the short-distance physics in a systematic way.  Most of that paper is in the context of an approximation 
 (their Eq.~2.24) that essentially assumes that the fractal dimension is larger than one.  
 The small scale structure was treated by an Ansatz~(their 3.16) which is equivalent to assuming that $\chi = \frac{1}{2}$.
 
 \subsection{Very Small Scales}
 
 At very small scales, we must take account of gravitational radiation, and also the matter-radiation transition.  According to Eq.~(\ref{physl}), the physical length of a segment grows as $t^{r(1-2\bar v^2)}$, or $t^{0.20}$ in the matter era.  Thus, a segment for which $l/t \sim 1$ at the time of matter-radiation equality will at later times have length $l_{\rm c}(t) \sim t(t/t_{\rm eq})^{r(1-2\bar v^2) - 1} \sim  t(t/t_{\rm eq})^{-0.80}$.  Segments larger than this will have emerged from the Hubble scale during the matter era, and so their small-scale structure will simply be of the matter-dominated form.  Segments smaller than this will have emerged during the radiation era, and so their structure will be of the radiation-dominated form with additional stretching during the matter era~\cite{Polchinski:2006ee}: 
 \begin{eqnarray} 
\left\langle [{\bf w}_+ (\sigma) - {\bf w}_+(\sigma')]^2 \right\rangle
&= &
{\cal A}_{\rm m} (l/t)^{2\chi_{\rm m}} \ ,\quad l > l_{\rm c}(t) \ , \\[3pt]
&=& {\cal A}_{\rm r}  (l_{\rm c}(t)/t)^{2(\chi_{\rm m} - \chi_{\rm r})} (l/t)^{2\chi_{\rm r}} 
\equiv {\cal A}_{\rm c}(t) (l/t)^{2\chi_{\rm r}}\ ,\quad l < l_{\rm c}(t)
  \ . \quad
\label{wwc}
\end{eqnarray}
In particular, the present value of ${\cal A}_{\rm c}$ is approximately $0.03$.

Now we can put in the smoothing via gravitational radiation, as discussed in Sec.~2.4.  The details details of the derivation~\cite{Siemens:2001dx,Siemens:2002dj,Polchinski:2007rg} are too lengthy for the present discussion, but we can summarize the result in a simple way.  Denote the Fourier transform of the two-point function~(\ref{fluct}$\,$,\ref{wwc}) by $\langle \tilde w^2(k) \rangle$.  Then, up to numerical constants, the string becomes smooth below a scale
\begin{equation}
l_{\rm GW} \sim \Gamma G\mu t \langle \tilde w^2(k_b)  \rangle\ , \quad k_b = 1/G\mu t\ .
\end{equation}
For the relevant tensions, the crossover form~(\ref{wwc}) is the relevant one, and
\begin{equation}
l_{\rm GW} \sim \Gamma {\cal A}_{\rm c}(t) (G\mu)^{1 + 2\chi_{\rm r}} t\ .  \label{lgw}
\end{equation}

\subsection{Loop Production}

We now take account of the production of small loops.  Since this takes place on scales small compared to the Hubble scale, we can go to locally flat null coordinates $u,v$, where ${\bf p}_+(u) = \partial_u {\bf x}$ and ${\bf p}_-(v) = \partial_v {\bf x}$, with $u + v = t$.

A loop forms whenever a long string passes through itself, meaning that  $ {\bf x}(u,v+l) =  {\bf x}(u+l,v)$ for two points on the string.  Defining
\begin{equation}
{\bf L}_+(u,l) = \int_u^{u+l} du \, {\bf p}_+(u)\ ,\quad
{\bf L}_-(v,l) = \int_v^{v+l} dv \, {\bf p}_-(v)\ ,
\end{equation}
this means that ${\bf L}_+(u,l) = {\bf L}_-(v,l)$ for some $u$, $v$, and $l$.  The rate of loop formation, per unit volume in $u, v, l$, is then
\begin{equation}
\langle \det J\, \delta^3( {\bf L}_+(u,l) - {\bf L}_-(v,l) ) \rangle\ ,\quad J = \frac{\partial^3 ( {\bf L}_+(u,l) - {\bf L}_-(v,l) )}{\partial u\, \partial v\, \partial l}\ .
\end{equation}
Now, the components of ${\bf L_{\pm}}$ are each proportional to the length $l$, so the $\delta$-function implies a factor of $l^{-3}$.  The rate would scale as $l^{-3}$ if the correlator of the $\bf w$'s were scale invariant.  To work more carefully, separate ${\bf p}_\pm$ as in Eq.~(\ref{ppw}), with the unit vectors ${\bf P}_{\pm}$ proportional to ${\bf L}_{\pm}$.  Then for small ${\bf w}_\pm$,
\begin{equation}
\delta^3( {\bf L}_+(u,l) - {\bf L}_-(v,l) )  = l^{-2} \delta^2({\bf P}_+ -{\bf P}_-) \delta(L_+(u,l) - L_-(v,l)) \ .
\end{equation}
The magnitudes $L_\pm$ are both equal to $l(1 - O(l^{2\chi}))$, so the whole is of order $l^{-3-2\chi}$.  The columns of $J$ are of order $l^\chi, l^\chi, l^{2\chi}$ (the last from the components parallel to ${\bf P}_\pm$), so the determinant is of order $l^{4\chi}$ giving $l^{-3 + 2\chi}$ for the rate.

The total rate of string length going into loops is weighted by an additional factor of $l$,
\begin{equation}
\int dl\, l^{-2 + 2\chi}\ . \label{loopdiv}
\end{equation}
This diverges at the lower end for $\chi \leq \frac{1}{2}$, a crucial result~\cite{Polchinski:2006ee}.  This large production of small loops seems surprising at first,\footnote{We thank Carlos Martins and Paul Shellard for suggesting this possibility.}  because the string is becoming smooth in the sense that the fractal dimension approaches 1.  What we have found in Eq.~(\ref{loopdiv}) is that the production of small loops is controlled by the rate at which the fractal dimension approaches~1, and in both the matter and radiation eras the approach is slower than the critical value~$\chi = \frac{1}{2}$.

The total rate~(\ref{loopdiv}) is determined by energy conservation, Eq.~(\ref{rholoop}).  Naively this requires a lower cutoff on the integral, around $l \sim 0.1 t$.  As we follow the evolution of the resulting loops, the same calculation leading to Eq.~(\ref{loopdiv}) shows that the production of small loops continues to diverge.  This suggests a complicated fragmentation process.  However, we have found a more physical way to introduce the cutoff~\cite{wip}.  If we separate the functions ${\bf p}_\pm$ into a `classical'  long-distance piece and a random short distance part (with our calculated two-point function), we find that loop production takes place only near intersections of the long-distance parts of ${\bf p}_+$ and ${\bf p}_-$ --- that is, near cusps (Fig.~3).
\begin{figure}
\begin{center}
\ \includegraphics[width=17pc]{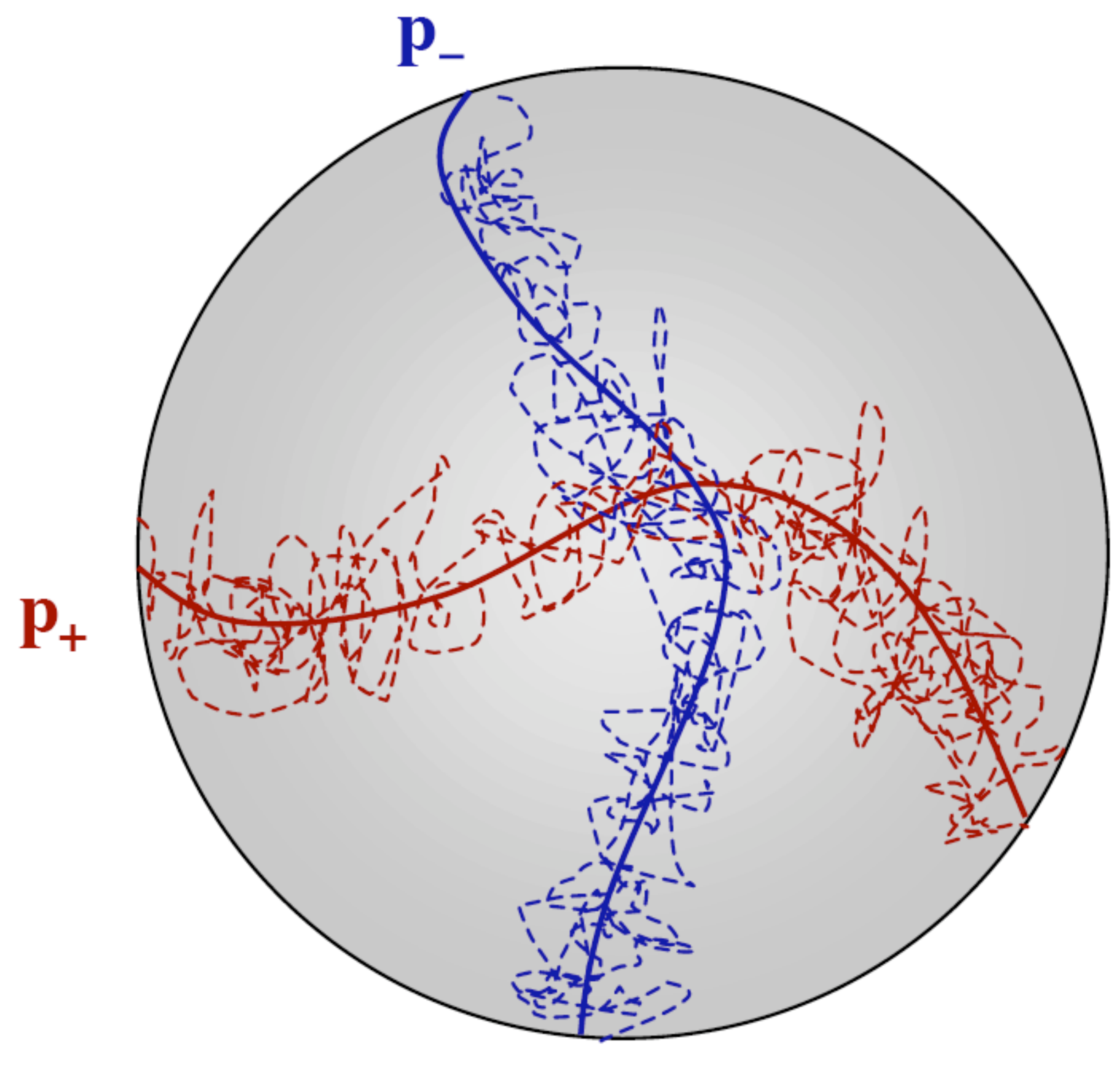}
\end{center}
\caption{The functions ${\bf p}_\pm$ written as a fixed long-distance piece plus a random short-distance part.  Production of small loops takes place near the cusp of the long-distance part.}
\end{figure}

Note that the figure is showing what is essentially the tangent vector to the string, rather than the string itself.  The latter has a fractal dimension that approaches 1 at short distance, but the tangent vector path has a large fractal dimension as indicated.  The two-point function~(\ref{fluct}) implies that a segment of length $\delta\sigma$ makes a step of order $\delta\sigma^{\chi}$.  After $N$ steps, the length of the path is $N \delta\sigma^{\chi}$, while the typical net distance covered~(\ref{fluct}) is $(N \delta\sigma)^{\chi}$, so the fractal dimension is $1/\chi$.  In particular, this implies that near each cusp of the long-distance curve, their are very many small cusps.  Formation of small loops occurs near these cusps: for a small loop, equality of ${\bf L}_+$ and ${\bf L}_-$ occurs when the vectors ${\bf p}_{\pm}$ are nearly parallel.

Away from the cusp, production of loops is shut down, but as the cusp is approached we find that production of loops starts uniformly on all scales down to the gravitational radiation cutoff, rather than fragmentating from large loops down to small.  This is still work in progress, but we appear to get a power law that extends all the way down to the gravitational radiation scale.

It should also be noted that these small loops are moving rather rapidly~\cite{Polchinski:2006ee}.  The velocity is just $L_+/l$, which by the above discussion is of order $1 - O({\cal A}[l/t]^{2\chi})$, so the relativistic $\gamma$ factor is of order ${\cal A}^{-1/2} (l/t)^{-\chi}$.  Recalling from footnote~\ref{lfoot} that $l$ is defined in terms of the energy of a string loop, the rest mass of a loop is of order $\mu l {\cal A}^{1/2} (l/t)^{\chi}$.

\sect{Conclusions}

\subsection{A Synthesis?}

Some of the most recent simulations~\cite{Vanchurin:2005yb,Vanchurin:2005pa,Olum:2006ix} have used a new technique to reach larger expansion factors than previously possible.  At intervals, the configuration is simply copied eight times, to produce a periodic box of twice the size.  A randomness is then briefly introduced into the evolution, to eliminate the unwanted long-distance correlations.  Various tests appear to indicate that no significant artifacts are introduced by this process.  

These simulations show two peaks in the loop production.  One is at a scale of order $0.1 t$, while the other remains near the short-distance cutoff.  In Refs.~\cite{Vanchurin:2005pa,Olum:2006ix} it is argued that the smaller peak is an artifact of the initial conditions and will eventually disappear.  However, it is this peak that we have found in the calculation of the previous section: these
small loops are an inevitable result of the small scale structure on the long strings.  We will be able to check this interpretation further when we complete the calculation of the small loop production function as described above.  The distribution functions shown in Ref.~\cite{Olum:2006ix} for the matter and radiation eras appear to match our Eq.~(\ref{loopdiv}) quite well, but a more careful study of loop fragmentation is needed~\cite{wip}.

We have not found the large loops, but it is inevitable that some of these will arise from chance self-intersections of long strings.  Our picture does explain why these large loops can survive fragmentation:
one would expect from Fig.~3 that the regions near cusps would fragment extensively, but non-cusp regions of the long loop will remain.  We must, at least for now, leave it to the simulations to determine the relative amount of string going into large and small loops.  From Ref.~\cite{Olum:2006ix}  it appears that only 20\% of the string goes into large loops during the matter era, and only 10\% during the radiation era.  However this reference also finds that the peak at large sizes is still growing at the end of the run, so the final numbers may be different.

Thus it appears that we may be approaching an accord on the question of loop sizes.  Some fraction of the string, perhaps a small fraction, goes into the largest scale under discussion, near the horizon.  The rest goes into loops at the gravitational radiation scale,  $l_{\rm f} \sim {\cal A}_{\rm c}(t) (G\mu)^{1 + 2\chi_{\rm r}} t$.  The simulations played an essential role in first revealing this double-peaked distribution, while the model plays a key role in confirming that the small loops, and in showing that gravitational radiation sets their actual size.

The divergence~(\ref{loopdiv}) explains why there is not just a single scale in the string network.  Indeed, the nonlinearities of string evolution have a remarkable ability to transfer energy from cosmic scales down to much shorter wavelengths.  In this respect the cusps are like shock waves.  Indeed, in shock waves the low energy field equations do break down, and shorter scale physics enters.  This has also been suggested for string networks~\cite{Vincent:1996rb,Vincent:1997cx,Bevis:2007gh}.  We believe that smoothing due to gravitational radiation prevents this, but cosmic string networks have been a source of many surprises, and perhaps more are in store.

\subsection{Gravitational Wave Signatures and the Inverse Problem}

Now let us consider the signatures, assuming that the picture reached above is correct: most of the string length goes into loops near the gravitational radiation scale, and a fraction (10\% in the radiation era and 20\% in the matter era) goes into loops of size $0.1 t$.\footnote{Ref.~\cite{Hogan:2006we}
gives a more complete treatment of future observations, including possible backgrounds.  Note however that it is based on a scenario where all of the string goes into loops of size $0.1 t$, and so the actual signatures will be reduced by a factor of 10 according to our picture.  Ref.~\cite{Siemens:2006yp} discusses gravitional waves from both large and small loops under varyious network scenarios.}  At pulsar frequencies, only the large loops are visible, so the energy density calculated in Sec.~3.4 is reduced by a factor of 10.  The bound~(\ref{pulbound}) is therefore weakened by a factor of 100, to $G\mu \lsim 2 \times 10^{-7}$, which is right around the CMB bounds (we have used the fact that these large loops are produced with relatively small velocities).  

In quoting such a ``bound" it is important to note that the theoretical uncertainty in the network properties is not included, because this is not known.  As one example, simulations of the full Abelian Higgs model seem to give long string densities that are a factor of three smaller~\cite{Bevis:2007gh} than simulations of the Nambu action, for reasons that are not understood.  Including this uncertainty weakens the CMB bounds by a factor of 3, but the pulsar bounds by a factor of 9 because $\Omega_{\rm GW}$ scales as $(G\mu)^{1/2}$.  The pulsar bounds have a further uncertainty coming from the sizes and numbers of the large loops.

The pulsar bound on $\Omega_{\rm GW}$ is expected improve by an order of magnitude from further observations at the Parkes Pulsar Timing Array~\cite{jenet}, improving the sensitivity to $G\mu$ by two orders of magnitude.   The future Square Kilometer Array is expected to reach several orders of magnitude further in $\Omega_{\rm GW}$, but at some point a stochastic background from 
coalescing black holes will dominate the string signal.   For the interferometers, Advanced LIGO sensitivity of $d\Omega_{\rm GW}/d\ln\nu_0 \sim 10^{-8}$ to $10^{-9}$ implies sensitivity to $G\mu$ down to $10^{-8}$ to $10^{-10}$. 

Turning to the high harmonics of the large loops, Fig.~3 makes it clear that the large loops form without large cusps: the world-sheet region near the intersection of the classical ${\bf p}_\pm$ curves fragments into small loops.  The functions ${\bf p}_\pm$ have large discontinuities where the cusps would be.  However, as the loops begin to decay via gravitational radiation, these curves smooth, because the higher harmonics decay faster than the lower.  Thus the large loops will develop cusps as they decay.  However, a factor of 10 is lost to the small loops (since large loops decaying today formed during the radiation era) so these are unlikely to be seen at LIGO except in very favorable parts of parameter space (such as small $P$).

The small loops form at the scale set by the gravitational smoothing.  We will adopt here our model from Sec.~4 to obtain representative numbers, but one should be mindful of the fact that the exponents and scales  that we have found may be corrected by future work; we do not have a systematic way to estimate errors.
For the relevant parameter range we will only see loops that form after the radiation era, so the relevant scale is Eq.~(\ref{lgw}), $l_{\rm f} \sim \Gamma {\cal A}_{\rm c}(t) (G\mu)^{1 + 2\chi_{\rm r}} t$.   However, because the loops are highly relativistic, their signatures are modified~\cite{Polchinski:2007rg}.  Thus, the formation scale $l_{\rm f}$ is $\mu^{-1}$ times the typical energy of a loop, giving
\begin{equation}
\epsilon_{\rm energy\mbox{\scriptsize -}length} \sim {\cal A}_{\rm c}(t) (G\mu)^{2\chi_{\rm r}}\ .
\end{equation}
This sets the period of oscillation as measured in the global FRW time,

However, most of the emitted radiation is Doppler-shifted by a factor $(1-v)^{-1} \sim \gamma^2 \sim {\cal A}_{\rm c}(t)^{-1} (l/t)^{-2\chi_{\rm r}}$.  For stochastic radiation the effective value of $\epsilon$ is set by the period of this radiation, and so
\begin{equation}
\epsilon_{\rm stoch} \sim   {\cal A}_{\rm c}(t)^{1 + 2\chi_{\rm r}} \Gamma^{2\chi_{\rm r}}   (G\mu)^{4\chi_{\rm r} (1 + \chi_{\rm r} )} \sim 0.03 (G\mu)^{0.44}\ .
\end{equation}
The frequency bound~(\ref{mingmu}) becomes
\begin{equation}
(G\mu)^{1.44} \gsim \frac{ 60 }{\Gamma  \nu_0 t_0}\ . 
\end{equation}
In the pulsar range this requires $G\mu \gsim 3 \times 10^{-7}$, so these small loops will not be seen, but in the LISA range it is sufficient that $G\mu \gsim  10^{-11}$, and in the LIGO range $G\mu \gsim 2 \times 10^{-14}$.  The density~(\ref{hiharm2}) becomes
\begin{equation}
\nu_0 \frac{d\Omega_{\rm GW}}{d\nu_0} = 10 c 
(\nu_0 t_0)^{-1/3} (G\mu)^{0.52} \ .  \label{hstoch}
\end{equation}
In the LIGO band this is of order $3 \times10^{-6} c (G\mu)^{0.52}$, and in the LISA band it is of order
 $10^{-4} c (G\mu)^{0.52}$\ .  Advanced LIGO sensitivity of $d\Omega_{\rm GW}/d\ln\nu_0 \sim 10^{-8}$ to $10^{-9}$ implies sensitivity to $G\mu$ only down to $10^{-5}$ to $10^{-7}$.  Potential LISA sensitivity down to $d\Omega_{\rm GW}/d\ln\nu_0 \sim 10^{-12}$ would allow detection of tensions down to $10^{-16}$, but the frequency limit of $10^{-11}$ is encountered first.
 
 For the signal of individual cusps, on the other hand, the effect of the boost goes in the opposite direction: the effective size of the cusp is increased by a factor of $\gamma$,
 \begin{equation}
 \epsilon_{\rm cusp\;size} \sim {\cal A}_{\rm c}(t)^{1/2 - \chi_r} \Gamma^{-2\chi_r}
 (G\mu)^{\chi_{\rm r} - 2\chi_{\rm r}^2}\ .
 \end{equation}
 This is around 0.02 over the range of interest.  The effect of $\epsilon_{\rm cusp} \sim 0.02$ is to increase by a factor $O(5)$ the event rate for $G\mu$ greater than $10^{-9}$, and decrease it by a factor $O(2)$ below this tension~\cite{Damour:2004kw}.  However, the relativistic loop velocities lead to two additional effects.  First, the number of cusp events is controlled by $\epsilon_{\rm energy\mbox{\scriptsize -}length}$, which is smaller by a factor of order 50; the rate is increased by this factor.  Second, the observed periodicity of a loop is set by $\epsilon_{\rm stoch}$: for $G\mu = 10^{-7}$ the periodicity is one year, but for $G\mu = 10^{-8}$ it drops to two weeks.  A one-year run will therefore see 25 cusps if it sees one, and so the mean event rate is not a good measure of the likelihood of seeing at least one event.\footnote{There are other such statistical effects.  For example, as Fig.~3 suggests, production of small loops is likely to be highly correlated in spacetime.  Ultimately such effects may lead to improved sensitivity, by fitting to templates based on rare events.  Ref.~\cite{Dubath:2007wu} examines one example of this type, the search for periodic signals from very small loops.}  Superimposing these effects on the results of Refs.~\cite{Damour:2001bk,Damour:2004kw,Siemens:2006vk} suggests that these small-loop cusps are within reach of Advanced LIGO only for tensions very close to the current upper limit, but again it is clear that a more precise understanding of the networks is essential. 
 
 Thus, of the four cases (large loops versus small and low harmonics versus high), the low harmonic stochastic radiation from large loops seems to be the most promising signature, with sensitivity to $G\mu$ improving by several orders of magnitude in the next decade, both in pulsar observations and at Advanced LIGO.  Assuming a vanilla string network, observation of the stochastic radiation would determine the string tension via~(\ref{stoch}), {\it assuming} that the large loop size and velocity distribution is accurately known, in particular the mean values of $\alpha^{1/2} \gamma^{-3/2}$ and the fraction of string going into long loops.  
 
In the approximation used in Sec.~3.4.1, the gravitational wave spectrum $\nu_0{d\Omega_{\rm GW}}/{d\nu_0}$ is nearly flat.  However, for $G\mu \lsim 10^{-8.5}$, the pulsar observations begin to see loops decaying during the matter era (see the discussion below Eq.~(\ref{nu0})), and so the energy density is larger, until $G\mu \sim 10^{-11}$ when the decaying loops are so small that their frequency is above the pulsar range.  The slope of the spectrum (and, ideally, the observation of the stochastic radiation in both the pulsar band and the LIGO band) give some check on the cosmic string interpretation.
 
It is interesting to go beyond the vanilla model and allow for a general value of $P$.  The simplest argument would predict that the signals scale as $1/P$, because one would need to increase the long string density by this factor in order to get the same intercommutation rate.  However, inserting a factor of $P$ into the loop production in the one-scale model~(\ref{1sm}) one finds that the model predicts that the density scales as $1/P^2$.  This is because one needs to {\it increase} the rate of long string intercommutations, to produces more kinks, so that the loop production (which only comes from self-intersections) stays at the $P=1$ value.  On the other hand, simulations~\cite{Sakellariadou:2004wq,Avgoustidis:2005nv} show a dependence more like $P^{-0.6}$: evidently, the reduction of $P$ is partly offset by multiple crossings of the colliding strings.  

There is some degeneracy in the determination of $G\mu$ and $P$.  Note that the low frequency stochastic radiation~(\ref{stoch}) and the high frequency stochastic radiation~(\ref{hstoch}) depend on $G\mu$ approximately as $(G\mu)^{0.5}$, and so on $G\mu$ and $P$ only through $(G\mu)^{0.5} P^{-0.6}$.  Thus, measurement of the overall normalization of the stochastic radiation does not break the degeneracy between $P$ and $G\mu$ even if both the low harmonic and high harmonic contributions are seen (unless $G\mu \lsim 10^{-8.5}$, as discussed above).
 
Ultimately, LISA will be sensitive to low-harmonic gravitational radiation from large loops, and individual cusps from large and small loops, over the most or all of the range of tensions that arise in brane inflation.  This will allow several independent determinations of $G\mu$ and $P$, making it possible to test the cosmic string hypothesis, and look for evidence of a less vanilla type of cosmic string.

\section*{Acknowledgments}
I would like to thank Florian Dubath and Jorge Rocha for discussions and collaboration.  I also thank Lars Bildsten, Ed Copeland, Thibault Damour, Gabriele Gonzalez, Mark Hindmarsh, Mark Jackson, Renata Kallosh, Andrea Lommen, Vuk Mandic, Carlos Martins, Ken Olum, Paul Shellard, Xavi Siemens, George Smoot, Tommaso Treu, and Alex Vilenkin for discussions and communications.  This work was  supported by NSF grants PHY05-51164 and PHY04-56556.

\vfill
\eject

\end{document}